\documentclass[aip, apl, reprint, floatfix]{revtex4-1}

\usepackage[T1]{fontenc}
\usepackage{amsfonts, amssymb, amsbsy, amsmath}
\usepackage{graphicx}
\usepackage{hyperref}
\usepackage{bm}
\usepackage{dcolumn}

\newcommand{\equref}[1]{(\ref{#1})}
\newcommand{\mr}[1]{\mathrm{#1}}
\newcommand{\be}{\begin{equation}}
\newcommand{\ee}{\end{equation}}

\newcommand{\figta}{$\left(\mathrm{a}\right)\;$}
\newcommand{\figtb}{$\left(\mathrm{b}\right)\;$}
\newcommand{\figtc}{$\left(\mathrm{c}\right)\;$}
\newcommand{\figtd}{$\left(\mathrm{d}\right)\;$}
\newcommand{\figte}{$\left(\mathrm{e}\right)\;$}

\newcommand{\figa}{$\left(\mathrm{a}\right)$}
\newcommand{\figb}{$\left(\mathrm{b}\right)$}
\newcommand{\figc}{$\left(\mathrm{c}\right)$}
\newcommand{\figd}{$\left(\mathrm{d}\right)$}
\newcommand{\fige}{$\left(\mathrm{e}\right)$}

\newcommand{\mhz}{\;\mr{MHz}}
\newcommand{\ghz}{\;\mr{GHz}}

\newcommand{\mk}{\;\mr{mK}}

\newcommand{\nh}{\;\mr{nH}}

\newcommand{\mm}{\;\mr{mm}}

\newcommand{\mum}{\;\mu\mr{m}}

\newcommand{\nm}{\;\mr{nm}}

\newcommand{\bext}{B_{\mr{ext}}}
\newcommand{\phiext}{\Phi_{\mr{ext}}}

\newcommand{\lk}{L_\mr{k}}

\newcommand{\phio}{\Phi_0}

\newcommand{\phie}{\varphi_{\mr{ext}}}

\newcommand{\ic}{I_\mr{c}}
\newcommand{\ej}{E_\mr{J}}
\newcommand{\ec}{E_\mr{C}}

\newcommand{\fq}{f_{\mr{q}}}
\newcommand{\fp}{f_{\mr{p}}}
\newcommand{\fs}{f_{\mr{s}}}

\newcommand{\el}{E_\mr{L}}

\newcommand{\ip}{I_\mr{p}}

\newcommand{\tc}{T_{\mr{c}}}
\newcommand{\dphi}{\delta\Phi}
\newcommand{\clen}{C_l}
\newcommand{\llen}{L_l}
\newcommand{\vph}{v}

\newcommand{\ql}{Q_{\mr{L}}}

\newcommand{\lsq}{L_{\square}}

\begin{document}

\title{Hybrid rf SQUID qubit based on high kinetic inductance}

\author{J. T. Peltonen}
\email{joonas.peltonen@riken.jp}
\altaffiliation[Present Address: ]{Low Temperature Laboratory, Department of Applied Physics, Aalto University School of Science, POB 13500, FI-00076 AALTO, Finland}
\affiliation{RIKEN Center for Emergent Matter Science, Wako, Saitama 351-0198, Japan}

\author{P. C. J. J. Coumou}
\affiliation{Kavli Institute of Nanoscience, Delft University of Technology, Lorentzweg 1, 2628 CJ Delft, The Netherlands}

\author{Z. H. Peng}
\affiliation{Key Laboratory of Low-Dimensional Quantum Structures and Quantum Control of Ministry of Education, Department of Physics and Synergetic Innovation Center for Quantum Effects and Applications, Hunan Normal University, Changsha 410081, China}
\affiliation{RIKEN Center for Emergent Matter Science, Wako, Saitama 351-0198, Japan}

\author{T. M. Klapwijk}
\affiliation{Kavli Institute of Nanoscience, Delft University of Technology, Lorentzweg 1, 2628 CJ Delft, The Netherlands}
\affiliation{Physics Department, Moscow State Pedagogical University, Moscow 119435, Russia}

\author{J. S. Tsai}
\affiliation{RIKEN Center for Emergent Matter Science, Wako, Saitama 351-0198, Japan}
\affiliation{Department of Physics, Tokyo University of Science, Kagurazaka, Tokyo 162-8601, Japan}

\author{O. V. Astafiev}
\email{Oleg.Astafiev@rhul.ac.uk}
\affiliation{Royal Holloway, University of London, Egham, Surrey TW20 0EX, United Kingdom}
\affiliation{National Physical Laboratory, Hampton Road, Teddington TW11 0LW, UK}
\affiliation{RIKEN Center for Emergent Matter Science, Wako, Saitama 351-0198, Japan}
\affiliation{Moscow Institute of Physics and Technology, 141700, Dolgoprudny, Moscow Region, Russia}

\begin{abstract}
We report development and microwave characterization of rf SQUID (Superconducting QUantum Interference Device) qubits, consisting of an aluminium-based Josephson junction embedded in a superconducting loop patterned from a thin film of TiN with high kinetic inductance. Here we demonstrate that the systems can offer small physical size, high anharmonicity, and small scatter of device parameters. The hybrid devices can be utilized as tools to shed further light onto the origin of film dissipation and decoherence in phase-slip nanowire qubits, patterned entirely from disordered superconducting films.
\end{abstract}

\date{\today}

\maketitle

\section{Introduction}

Various applications of superconducting quantum bits (qubits), see for example Refs.~\onlinecite{devoret13} and~\onlinecite{barends14}, benefit from building blocks with good reproducibility of device parameters, high anharmonicity of the energy level spacings, and compact physical size. These requirements apply in particular to the case of superconducting quantum metamaterials~\cite{zagoskin14,macha14}, where a large number of identical or controllably different ``artificial atoms'' are required. In typical flux qubits~\cite{chiorescu03} based on three or four Josephson tunnel junctions (JJs) one of the most significant issues is the exponential sensitivity of the transition frequency on the potential barrier height and hence the precise tunnel junction geometry and transparency. Optimized device design and fabrication process~\cite{yan16,orgiazzi16} can mitigate this effect along with the steepness of the energy bands and poor decoherence properties away from the optimal flux working point. Promising decoherence times and large anharmonicities have been predicted for inductively shunted JJs~\cite{koch09,zorin09}. They have been realized also experimentally~\cite{manucharyan09,vool14,pop14}, in particular in the fluxonium configuration~\cite{manucharyan09,rastelli15,viola15}, where a single phase-slip junction closes a superconducting loop with high inductance, typically formed by a long series array of larger JJs~\cite{masluk12,bell12}.

\begin{figure}[htb]
\includegraphics[width=\columnwidth]{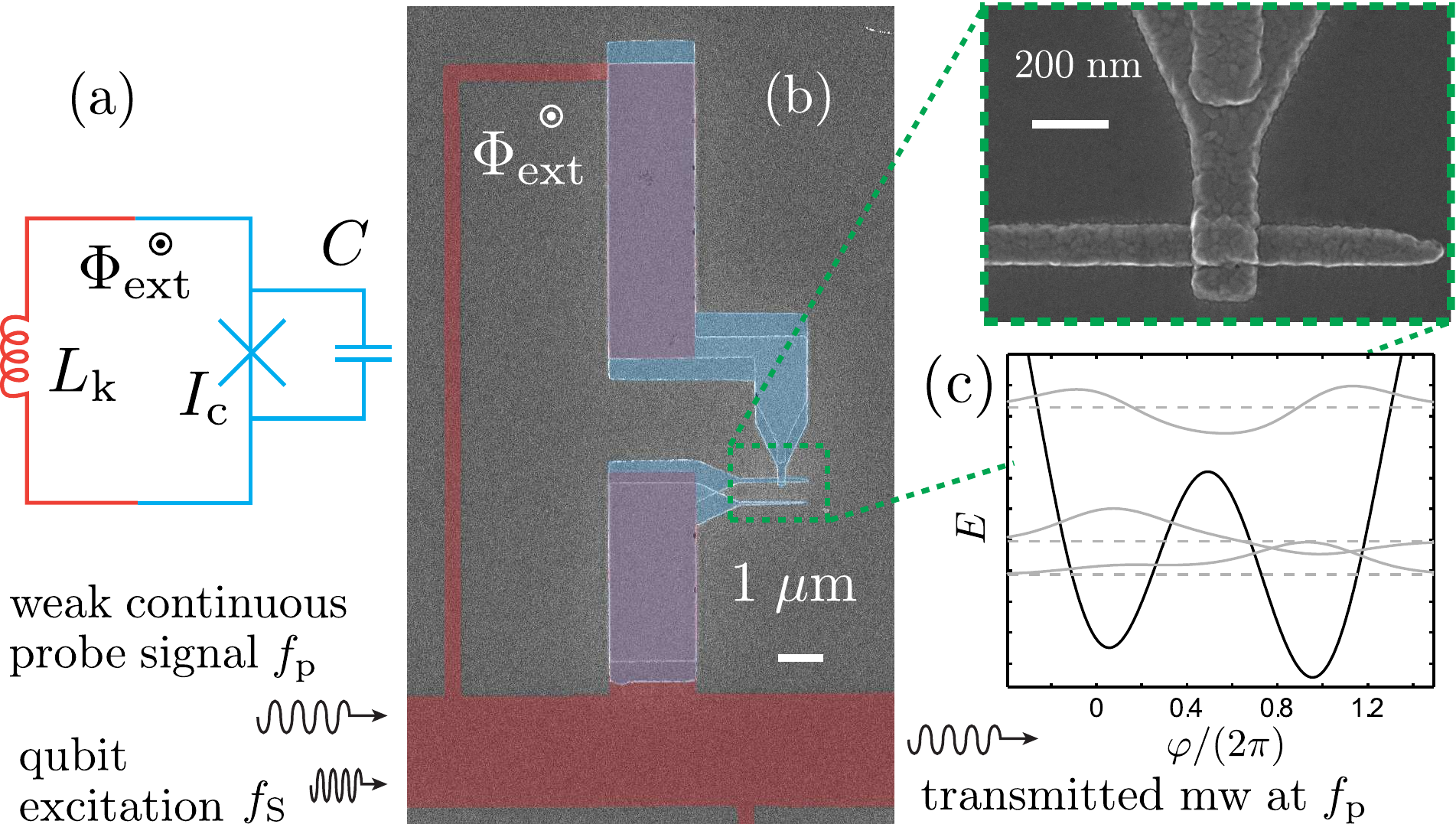}
\caption{\figta Schematic circuit representation of a hybrid rf SQUID. A superconducting loop with high kinetic inductance (red) is closed with a single Josephson junction (blue) and placed into perpendicular external magnetic field. \figtb False-color scanning electron micrograph of TiN--Al rf SQUID investigated in this work. The TiN loop is shaded in red, whereas the Al-based tunnel junction is shown in blue, and the direct galvanic contact overlap areas in purple. \figtc Sketch of the potential $U(\varphi)$ (black solid line) for $\phiext/\phio=0.56$, together with the three lowest-lying energy levels (horizontal gray dashed lines) and the corresponding wavefunctions (gray solid lines) from the rf SQUID Hamiltonian for parameters typical to the measured devices.} \label{fig:sample}
\end{figure}

In this work, we develop and present an experimental study towards flux qubits in the basic rf SQUID geometry of a single Josephson junction shunted by the inductance of a superconducting loop~\cite{wendin06,friedman00,likharevbook}. Crucially, in our devices the loop inductance is dominated by the kinetic inductance of a thin disordered superconducting film, {\it cf}. Fig.~\ref{fig:sample}. Analogously to the use of a junction array to form the highly inductive environment for the active qubit junction, this approach allows to realize a large loop inductance in compact size. Our motivation for the study of the system is threefold: First, we seek to demonstrate such a hybrid superconducting quantum system, and to investigate the feasibility of this simple archetype of a flux qubit. Secondly, we look to employ the hybrid structure, combining a standard aluminium-based JJ with the loop made of an ultrathin superconductor close to the superconductor-to-insulator transition, as a tool to assess film-induced decoherence and dissipation in phase-slip nanowire qubits patterned entirely from such superconductors with high kinetic inductance~\cite{mooij05,mooij06,astafiev12,peltonen13}. Thirdly, our devices, featuring an inductively shunted JJ, pave the way for transport measurements of phase-slip physics in this basic system~\cite{guichard10,dimarco15}, complementing existing work on JJ arrays~\cite{weissl15,manucharyan12}.

Figure~\ref{fig:sample}~\figta shows a schematic of a hybrid rf SQUID of the above type: The superconducting loop has total kinetic inductance $\lk$, giving rise to the inductive energy scale $\el=\phio^2/(4\pi^2\lk)$. Likewise, the junction has critical current $\ic$ and capacitance $C$, resulting in the Josephson energy $\ej=\hbar\ic/2e$ and charging energy $\ec=e^2/2C$. The SQUID loop is placed in a perpendicular external magnetic field $\bext$, giving rise to the flux $\phiext$ threading the loop. Figure~\ref{fig:sample}~\figtc further shows a sketch of the SQUID double well potential $U(\varphi)=\ej(1-\cos\varphi)+\el(\varphi-\phie)^2/2$ (see, for example, Ref.~\onlinecite{wendin06}), as well as the three lowest energy levels and wave functions calculated for $\phie=2\pi\times 0.56$, and the representative parameters $\el\approx 4.5\ghz$, $\ej\approx 41\ghz$, and $\ec\approx 18\ghz$, yielding $\fq\approx 11.1\ghz$. These values corresponding to device I in Fig.~\ref{fig:spect}. Here, the control phase $\phie$ is related to the externally applied biasing magnetic flux $\phiext$ via $\phie=2\pi\phiext/\phio$.

\section{Sample details}

Fabrication of the hybrid structure is a technologically challenging problem. The key element is a galvanic contact between the thin film of the highly disordered material and Al. The false color scanning electron micrograph in Fig.~\ref{fig:sample}~\figtb illustrates a typical single rf SQUID studied in this work, together with a sketch of the measurement setup. The approximately $400\nm$ wide TiN wire that provides the kinetic inductance is shaded in red, whereas the Al-AlOx-Al JJ, fabricated by two-angle shadow evaporation and closing the TiN loop, is highlighted in blue. The two large TiN--Al contact overlap areas are colored purple. The bottom TiN loop edge doubles as part of the $2.5\mum$ wide resonator center line, widening to $5\mum$ outside the center section with the SQUID loops. This shared mutual kinetic inductance facilitates the inductive SQUID--resonator coupling.

To pattern inductances from the TiN films, we used a process similar to Refs.~\onlinecite{peltonen13} and~\onlinecite{peltonen16}, relying strongly on electron beam lithography (EBL). The starting point is an oxidized Si wafer onto which a thin film of TiN with thickness $d\approx 6\nm$ is grown by atomic layer deposition (ALD)~\cite{coumou13,coumou13b,bueno15}. This TiN film is identical to film A in Ref.~\onlinecite{coumou13b}. First, a mask for the CPW resonator ground planes [not visible in Fig.~\ref{fig:sample}~\figb] as well as coplanar transmission lines for connecting to the microwave measurement circuit is defined by EBL. These structures are consequently metallized in an electron gun evaporator with $5\nm$ Ti, $70\nm$ Au, and $10\nm$ Al on top. After liftoff, another layer of resist is applied by spin coating, and patterned in a second step of EBL to act as an etch mask for the TiN loops and the resonator center line, {\it i.e.}, the structures highlighted in red in Fig.~\ref{fig:sample}~\figb. The pattern is transferred into the TiN film by reactive ion etching (RIE) with $\mr{CF}_4$ plasma.

Following the etching step, the remaining resist is removed, and a new bilayer resist is applied to prepare for the last EBL step for defining the Josephson junction, blue in Fig.~\ref{fig:sample}~\figb, to close the TiN loop. After development, the mask is loaded into an UHV e-gun evaporator. Crucially, prior to Al deposition the exposed TiN contact surfaces, purple in in Fig.~\ref{fig:sample}~\figb, are cleaned by a brief {\it in-situ} Argon ion milling. Immediately after this, the typically $30\nm$ thick Al electrodes of the JJ are deposited by conventional shadow evaporation at two different tilt angles. The two Al depositions are separated by an {\it in-situ} oxidation in a 10\%--90\% mixture of $\mr{O}_2$ and Ar to form the AlOx tunnel barrier. To protect the TiN film from oxidation, the samples were stored under nitrogen atmosphere, and cooled down within 1--2 days after removing the protective resist. The resonator chip was enclosed in a sample box, and microwave characterization was performed in a dilution refrigerator at the base temperature close to $25\mk$. Samples from several fabrication rounds with differing Ar ion cleaning and oxidation parameters were cooled down. Here we present measurement results belonging to one typical sample.

From low temperature dc transport measurements of separate test structures, we infer sufficient quality of the TiN--Al contacts, supporting supercurrents $\gg\ic$, the critical current of the SQUID Al junction, and showing no significant suppression of the transition temperature $\tc$ of the TiN film due to the Ar ion cleaning. Similarly, suitable JJ oxidation parameters were determined by room temperature resistance measurements of a series of junctions with differing overlap areas.

\section{Microwave characterization}

To characterize the devices we use a vector network analyzer to monitor the transmission of microwaves through the resonator, at probing frequencies $\fp$ close to one of the resonant modes $f_n=n\vph/2L$, $n=1,2,3,\ldots$. Here, $L$ denotes the resonator length and $\vph=1/(\llen\clen)^{1/2}$ the effective speed of light, expressed in terms of $\llen$ ($\clen$), the inductance (capacitance) per unit length. The samples reported here contain a resonator with $L=1.5\mm$, resulting in the fundamental mode frequency $f_1~\approx 2.5\ghz$ with loaded quality factor $\ql\approx 1\times 10^3$.

\begin{figure}[htb]
\includegraphics[width=\columnwidth]{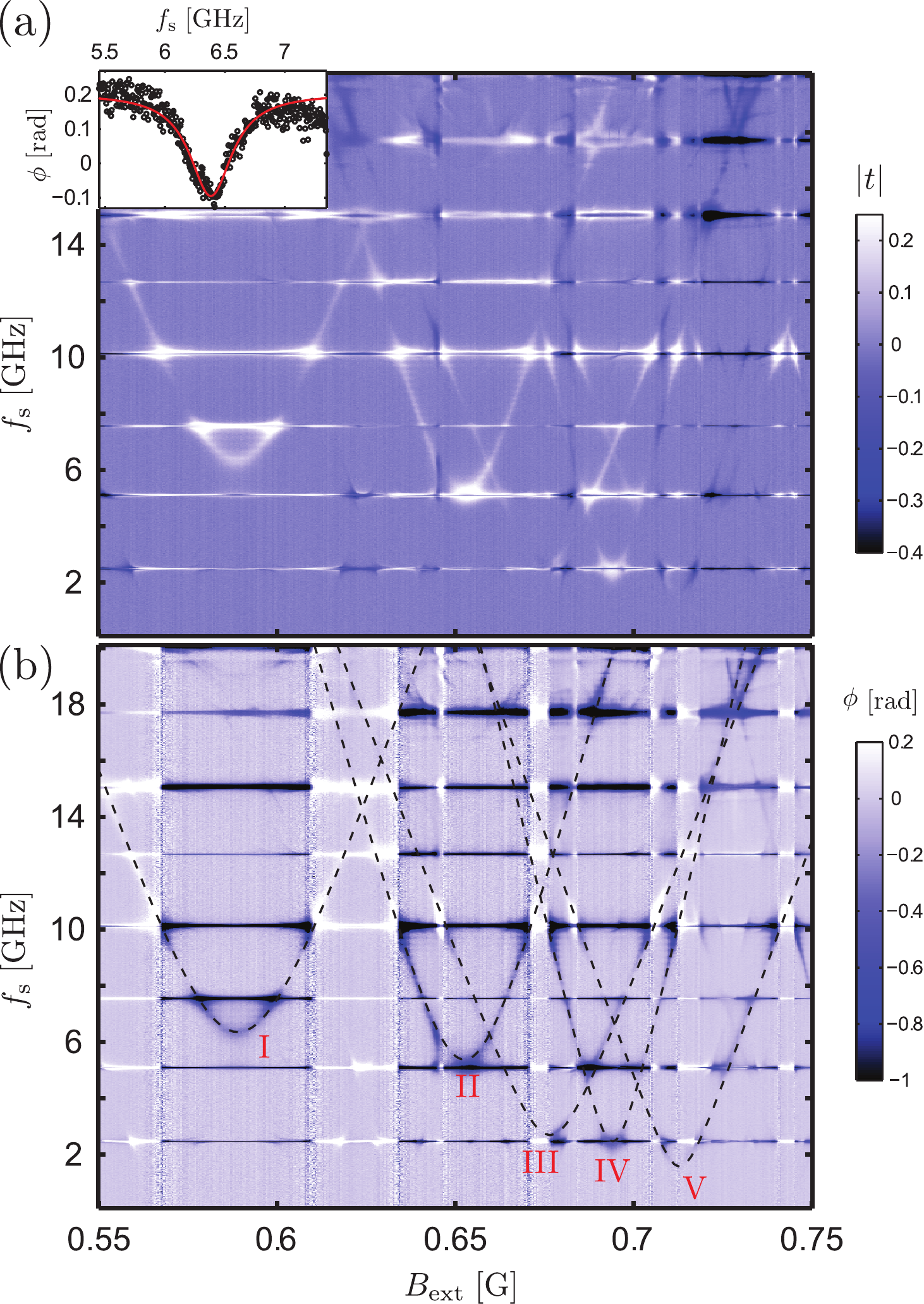}
\caption{\figta Typical two-tone spectroscopy in a narrow range of the external magnetic field $\bext$, showing the amplitude change of mw transmission through the resonator, probed at a fixed frequency at one of the resonant modes. The horizontal lines arise due to the multiple resonator modes. Inset: spectroscopy lineshape at the optimal point for the leftmost transition evident in the main panel (device I with $\Delta\approx 6.3\ghz$). \figtb The same spectroscopy measurement as in panel \figa, now showing the phase change of the mw transmission coefficient. The dashed lines correspond to theoretically calculated qubit frequencies $\fq$ vs. $\bext$ for five devices with the strongest signatures in this range of $\bext$.}
\label{fig:spect}
\end{figure}

Signatures from the SQUID loops become visible as the global external magnetic field $\bext$ is scanned. In a typical initial test this is done over a period corresponding to $\phiext$ of several flux quanta through the loops. At the input port of the resonator, the low-power probing tone at frequency $\fp$ is combined with another continuous microwave signal at frequency $\fs$ for exciting the qubits. A representative result of such two-tone spectroscopy is illustrated in the top panel of Fig.~\ref{fig:spect}, focused on a range of $\bext$ with transitions belonging to five loops coupled to the same resonator. In this measurement, showing the magnitude change of the transmission coefficient, the weak probe tone was fixed at $\fp=f_4$ while the frequency $\fs$ of the strong drive signal was scanned across a large span.

The bottom panel of Fig.~\ref{fig:spect} displays the corresponding phase change of $t$, together with dashed lines indicating qubit transition frequencies calculated according to the standard rf SQUID Hamiltonian~\cite{wendin06}
\be
H=\ec\hat{n}^2-\ej\cos\hat{\varphi}+\el(\hat{\varphi}-\phie)^2/2.\label{eq:hamiltonian}
\ee
They are obtained by finding the lowest energy eigenstates by exact diagonalization. In Eq.~\equref{eq:hamiltonian}, the number operator $\hat{n}$ of the charge on the junction capacitor and the phase operator $\hat{\varphi}$ obey the commutation relation $[\hat{\varphi},\hat{n}]=i$. Close to $\phiext=(N+1/2)\phio$, the shape of the curves is well approximated by $h\fq=\sqrt{\varepsilon^2+\Delta^2}$. Here $\varepsilon=2\ip\dphi$ with $\ip$ denotes the persistent current, and we introduced the flux deviation from degeneracy, $\dphi=\phiext-(N+1/2)\phio$. 

The inset of Fig.~\ref{fig:spect}~\figta further shows the spectroscopy signal lineshape for the SQUID with $\Delta/h\approx 6.3\ghz$ (device I the main plot), in the low power limit of the spectroscopy tone, together with a Lorentzian fit. For different devices, we find typical HWHM values between $50-300\mhz$ at the optimal point, depending on the detuning from the nearest resonator modes and transitions due to the other SQUID loops. We emphasize that this is the first study of the hybrid TiN--Al devices, and the coherence can be further improved by optimizing the geometry and improving the film quality.

\begin{figure}[htb]
\includegraphics[width=\columnwidth]{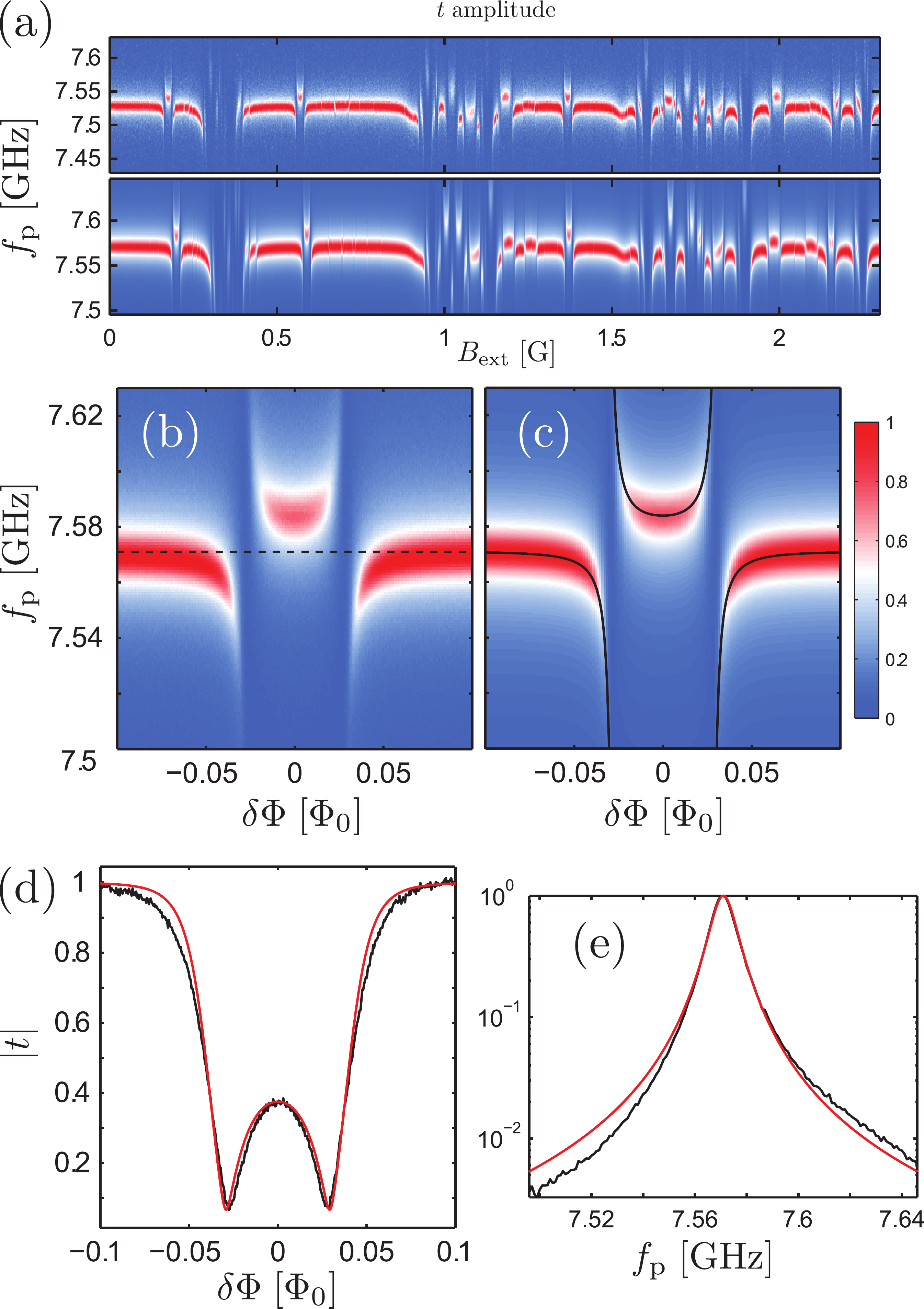}
\caption{\figta Normalized mw transmission coefficient amplitude $|t|$ for two nominally identical samples, fabricated simultaneously and characterized in the same cooldown cycle. After detailed analysis, fingerprints from 23 out of the 30 SQUID loops can be distinguished. \figtb Measured features in $|t|$ due to a single rf SQUID, compared to the calculated transmission in~\figc. Panel~\figtd shows a comparison of line cuts of \figtb and \figtc at constant $\fp=f_3$, indicated by the horizontal dashed line in \figb. In panel~\fige, the lineshape of the bare resonator mode (black) is compared with a Lorentzian fit (red).} \label{fig:fingerprints}
\end{figure}

Figure~\ref{fig:fingerprints}~\figta compares the $\bext$-dependent transmission amplitudes for $\fp$ around a narrow range centered at $f_3$. The two panels correspond to two nominally identical samples cooled down simultaneously, demonstrating good reliability of the TiN -- Al contacts and a promising degree of reproducibility. After detailed analysis of the periodicities of the various features, we detect fingerprints from 23 out of the total 30 SQUID loops, with the largest predicted values of $\Delta$. The remaining devices with $\Delta/h\ll 500\mhz$ are likely to be functional as well, although with too weak coupling for their features to be resolved in this measurement. The bottom panel corresponds to the sample in Fig.~\ref{fig:spect} as well as Fig.~\ref{fig:energies} below.

The behavior of $|t|$ at the individual anticrossings due to the qubit transitions can be modeled accurately using a model based on a standard Lindblad master equation~\cite{oelsner10,peltonen13}. In panel~\figtb of Fig.~\ref{fig:fingerprints} we show in an enlarged view the measured features in the normalized transmission amplitude $|t|$ due to the anticrossings of a single qubit (device I in Fig.~\ref{fig:spect}). The plot is a zoom-in to a short section of the data in the bottom panel of Fig.~\ref{fig:fingerprints}~\figa. Panel~\ref{fig:fingerprints}~\figtc displays the transmission amplitude calculated with the master equation-based model~\cite{oelsner10,peltonen13}, in good agreement with the measurement.

The black solid lines indicate the $\bext$-dependence of two of the eigenstates of the hybridized qubit--resonator system. The horizontal black dashed line shows the bare resonator frequency $f_3$, the value of $\fp$ at which the 1D line cuts of $|t|$ in Fig.~\ref{fig:fingerprints}~\figtd are plotted as a function of $\bext$. In panel~\figte we further plot the bare resonator transmission for $\fp$ around $f_3$ as the black solid line, at a constant $\bext$ when all the qubit transitions are well detuned from this resonator mode. The red line is a Lorentzian fit included for reference.

After comparisons [as in Fig.~\ref{fig:fingerprints}~\figb--\figd] of the transmission measurements with the theoretical model for several qubit transitions visible in both of the two resonators presented in Fig.~\ref{fig:fingerprints}~\figa, we can indirectly approximate the scatter in $\Delta$ to be less than 5 \% for the qubits with the largest $\Delta$. For the initial samples reported here all the SQUIDs had different parameters by design, mainly the combination of the loop length and junction size. In addition, the number of of well-isolated features is limited due to the large number of loops in each resonator. To get a more accurate estimate of the fabrication scatter in $\Delta$ and other device properties, future experiments will therefore investigate fewer nominally identical SQUIDs coupled to the same resonator, and include a detailed comparison of two nominally identical resonators.

\begin{figure}[htb]
\includegraphics[width=\columnwidth]{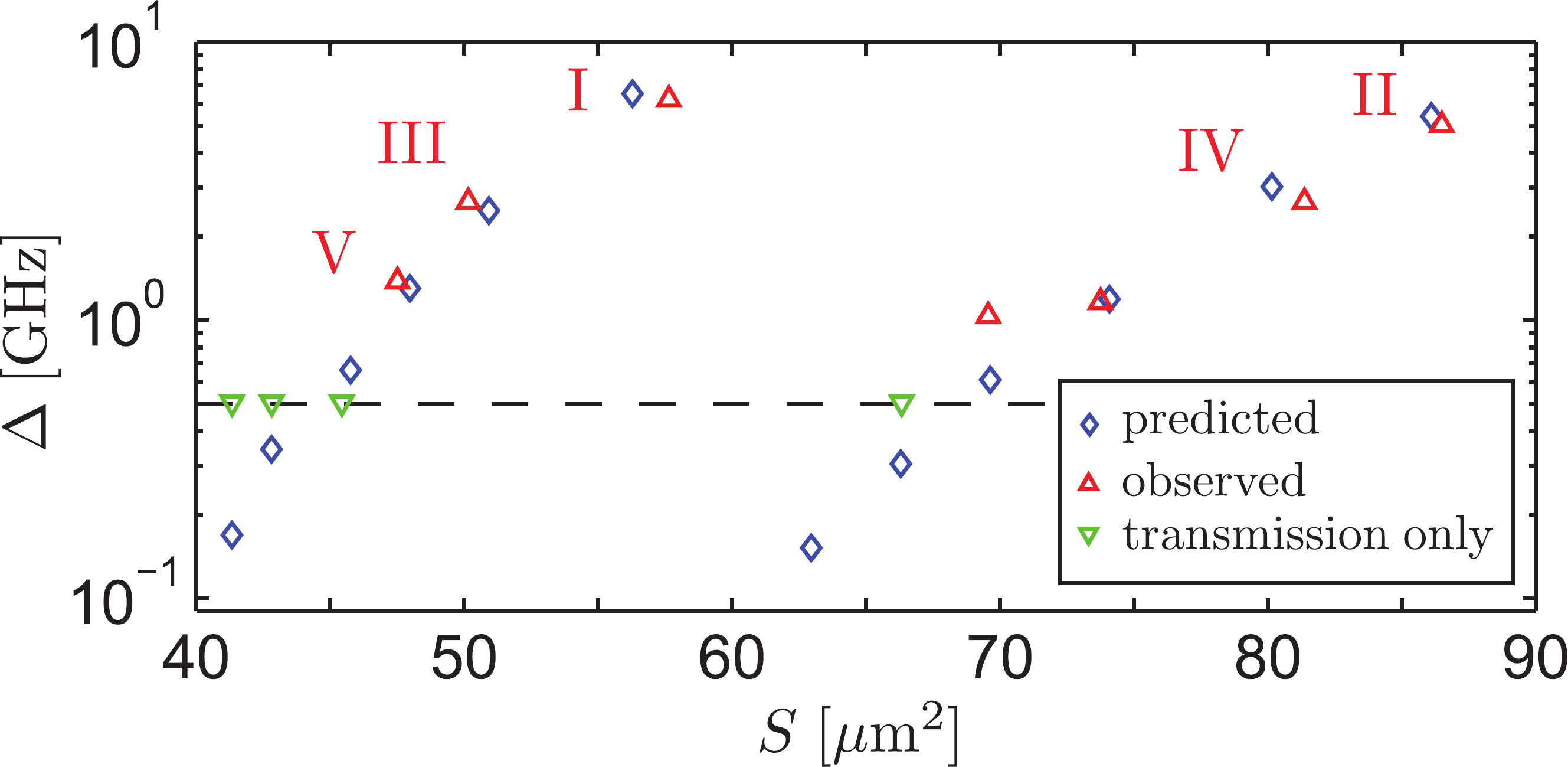}
\caption{Observed rf SQUID energy gaps $\Delta$ at the optimal points for one sample, extracted from two-tone spectroscopy measurements similar to Fig.~\ref{fig:spect}, and plotted against and effective loop area obtained from the the observed periodicities with magnetic field. The symbols $\nabla$ and $\triangle$ show the experimental points. They are compared with theoretical predictions ($\diamond$) based on the standard rf SQUID Hamiltonian (see text for details).} \label{fig:energies}
\end{figure}

In Fig.~\ref{fig:energies} we collect together the minimum qubit energy gaps $\Delta$ at the optimal flux points, for one of the measured chips. They are shown as the red upward-pointing triangles, extracted from fits to two-tone spectroscopy measurements similar to Fig.~\ref{fig:spect}. Our present scheme is sufficient for resolving qubits in two-tone spectroscopy if $\Delta/h\gtrsim 1\ghz$. Devices with $\Delta/h<1\ghz$ remain visible in direct transmission measurements, {\it cf.} Fig.~\ref{fig:fingerprints}. However, the exact value for $\Delta$ in this case can be only indirectly inferred from a comparison of the numerically simulated transmission coefficient with the measurement. For $\Delta/h\ll\fp$ this leads to a large uncertainty, and hence these devices with low $\Delta$ are indicated at $500\mhz$ (green down-triangles).

The experimental values of $\Delta$ are plotted against the effective loop area $S$, deduced from the $\bext$ -periodicity of the spectroscopy lines. Analogously, for the SQUIDs with the lowest $\Delta$, the values of $S$ were determined by the $\bext$ -periodicity of features in direct transmission measurements.

The sawtooth behavior evident in $\Delta$ vs. $S$ in Fig.~\ref{fig:energies} is due to the designed variation in the JJ width. On the other hand, the inductive energy $\el$ was designed to decrease monotonously with increasing loop area $S$, corresponding further to increasing $\lk$. To compare these observed energy gaps $\Delta$ with theoretical predictions, we use the rf SQUID Hamiltonian of Eq.~\ref{eq:hamiltonian}. As input parameters we use the sheet kinetic inductance $\lsq\approx 1.3\nh$ determined independently from the resonator properties, as well as the nominal loop areas and the number of squares of TiN in each of loops. In addition, we use JJ overlap areas obtained from SEM observations. They differ from the nominal design overlaps, by a approximately constant offsets of $50\nm$ and $30\nm$ in the width and the height of the junction, respectively. Then, using as adjustable parameters only the values $70\;\mr{fF}/\mu\mr{m}^2$ and $6\;\mu\mr{A}/\mu\mr{m}^2$ of the specific junction capacitance and critical current, respectively, we find reasonable agreement between the predictions of the model (blue diamonds) and the experimental observations. Notably, we assume the same values for these oxidation parameters for all the junctions.

\section{Conclusions}

In summary, we have developed and investigated properties of hybrid rf SQUID qubits relying on the high kinetic inductance of a thin, disordered superconducting film. We find reasonable reproducibility of the device parameters. Future samples will benefit from having only one qubit coupled to a single, hanger-style resonator, several of which can be multiplexed to a single readout transmission line. We note that a somewhat thicker TiN film can be straightforwardly used for forming an equally large loop inductance, in the form of a meander. Then, making the contact is likely to be easier as well as the qubits to be subjected to less dissipation. Moreover, the Ar milling step can be further separately optimized.

Due to the robust fabrication process, the hybrid rf SQUIDs can be employed as a characterization tool and to provide a further control check of decoherence in phase-slip qubits, pointing towards film losses. The present work, demonstrating the ability to create good contact between the thin TiN film and subsequently evaporated Al structures, will be further relevant for dc transport measurements dealing with phase-slip physics of Josephson junctions in highly inductive environments.

\begin{acknowledgments}
We thank K. Kusuyama for assistance with sample fabrication. The work was financially supported by the JSPS FIRST program, and MEXT Kakenhi 'Quantum Cybernetics'. J. T. P. acknowledges support from Academy of Finland (Contract No. 275167). O. V. A. thanks Russian Science Foundation (grants N 15-12-30030 and N 16-12-00070) for support of this work. T. M. K. thanks the support by the Ministry of Education and Science of the Russian Federation, contract No. 14.B25.31.0007 of 26 June 2013, and by the European Research Council Advanced grant no. 339306 (METIQUM).
\end{acknowledgments}

\end{document}